\documentclass[conference,a4paper]{APSIPA2021}
\usepackage[dvips]{graphicx}
\usepackage{amsxtra}
\usepackage{threeparttable}
\usepackage{amsmath,amssymb}
\usepackage{enumerate}
\usepackage{comment}
\usepackage{float}
\usepackage{bm}
\usepackage{hhline,amsmath}
\usepackage{multirow}
\usepackage{algorithmic}
\usepackage{url}
\usepackage{bm}

\usepackage{colortbl}
\usepackage{array}
\usepackage{bbding}
\usepackage{pifont}
\usepackage{wasysym}
\usepackage{setspace}
\usepackage{chngpage}
\usepackage{hhline,amsmath}
\newlength\savedwidth
\newcommand{\wcline}[1]{\noalign{\global\savedwidth\arrayrulewidth\global\arrayrulewidth 1.0pt}
\cline{#1}
\noalign{\global\arrayrulewidth\savedwidth}}

\begin{document}

\title{How Should We Evaluate Synthesized\\ Environmental Sounds}

\author{%
\authorblockN{%
Yuki Okamoto\authorrefmark{1},
Keisuke Imoto\authorrefmark{2},
Shinnosuke Takamichi\authorrefmark{3},\\
Takahiro Fukumori\authorrefmark{1}, and
Yoichi Yamashita\authorrefmark{1}
}
\authorblockA{%
\authorrefmark{1}
Ritsumeikan University, Japan \\
}
\authorblockA{%
\authorrefmark{2}
Doshisha University, Japan\\
}
\authorblockA{%
\authorrefmark{3}
The University of Tokyo, Japan\\
}
}

\maketitle
\thispagestyle{empty}
\begin{abstract}
  Although several methods of environmental sound synthesis have been proposed, there has been no discussion on how synthesized environmental sounds should be evaluated.
  Only either subjective or objective evaluations have been conducted in conventional evaluations, and it is not clear what type of evaluation should be carried out.
  In this paper, we investigate how to evaluate synthesized environmental sounds.
  We also propose a subjective evaluation methodology to evaluate whether the synthesized sound appropriately represents the information input to the environmental sound synthesis system.
  In our experiments, we compare the proposed and conventional evaluation methods and show that the results of subjective evaluations tended to differ from those of objective evaluations.
  From these results, we conclude that it is necessary to conduct not only objective evaluation but also subjective evaluation.
\end{abstract}

\section{Introduction}
Environmental sounds are widely used in media content such as animated cartoons, video games, and movies.
Environmental sounds are important in directing situations and scenes in media content.
However, there is a limit to the amount of sound data that is openly available.
To obtain a variety of sounds, some statistical methods have been proposed to synthesize artificially environmental sounds \cite{okamoto_arXiv_2019_01,Kong_ICASSP2019_01,Liu_arXiv_2020_01,Yang_arXiv2022_01,Chen_IEEE_2020_01}.
Environmental sound synthesis has many potential applications such as the creation of background and sound effects for media content \cite{Lloyd_ACMI3DGG_01,Zhou_CVPR2018_01} and data augmentation for environmental sound analysis \cite{Kong_ICASSP2019_01,Wang_VR2017_01,Gontier_ICASSP_2020,Salamon_WASPAA2017_01}. 
It is also expected to produce realistic sound spaces in virtual spaces and further development of metaverse services.

To evaluate synthesized environmental sounds, some objective and subjective evaluation methods have been proposed \cite{okamoto_arXiv_2019_01,Liu_MLSP2021_01}.
Although many evaluation methods exist, there is still no established methodology.
In other words, it is not clear whether objective or subjective evaluation will be sufficient and what should be evaluated.
Conventionally, objective evaluation tends to be performed more often, owing to the low cost of evaluation.
However, there is no objective evaluation metric that fully corresponds to the subjective quality of synthesized sounds.
Indeed, the main evaluation of statistical speech synthesis competitions \cite{Zhou_BlizzardChallenge_2020} is conducted subjectively by human listeners.
We believe that subjective evaluation is also necessary for evaluating synthesized environmental sounds for human listening.
In addition, in conventional subjective evaluation \cite{Kong_ICASSP2019_01,Liu_MLSP2021_01}, the subjective quality of the synthesized sound is measured, and whether the sound reflects the input information used to synthesize it is not evaluated.

In this paper, we propose a methodology for the subjective evaluation of synthesized environmental sounds. 
Specifically, we propose a methodology to evaluate the degree of consistency between the synthesized environmental sound and the input information used to synthesize it.
The evaluation methodology is applied to the synthesis of environmental sounds from sound event labels, onomatopoeia, or both.
Furthermore, we demonstrate the necessity of subjective evaluation by comparing its result with that of objective evaluation.

The remainder of this paper is structured as follows.
In Sec. \ref{SecII}, we describe the conventional evaluation method for environmental sound synthesis.
In Sec. \ref{SecIII}, we present our proposed evaluation methodology for environmental sound synthesis.
In Sec. \ref{SecIV}, we discuss experiments where we compared the results of objective and subjective evaluations.
Finally, we summarize and conclude this paper in Sec. \ref{SecV}.

\section{Conventional evaluation method for environmental sound synthesis}
\label{SecII}
In this section, we introduce the existing methods of evaluating synthesized environmental sounds using objective and subjective modes.
\subsection{Conventional Objective Evaluation Method}
Liu et al. proposed an objective evaluation method using a sound event classifier for synthesized sounds from a sound event label \cite{Liu_MLSP2021_01}.
The classifier is trained using training data from environmental sound synthesis. 
They consider high-quality sounds to be classified into the correct class.
This objective evaluation method is useful for the system used for data augmentation of environmental sound analysis, for example, to help the computer distinguish sound characteristics.
However, the fact that the sound is accurately classified does not necessarily mean that humans perceive it to be of high quality. 
For example, environmental sounds used in media content must be sounds that humans perceive to be of high quality.
The conventional objective evaluation is insufficient, and subjective evaluation is still needed.
\begin{table*}[t!]
\footnotesize
\caption{List of synthesis methods for each evaluation method}
\vspace{2pt}
\label{table:synthesize_method}
\centering
\begin{tabular}{@{}lcc|cccc|c@{}}
    \wcline{1-8}
     &\\[-6pt]
     & \multicolumn{2}{c|}{\scalebox{0.98}{Input features}} & \multicolumn{4}{c|}{\scalebox{0.98}{Subjective evaluation method}} & \multirow{2}{*}{\scalebox{0.98}{Objective evaluation}}\\
     \cline{2-7}\\[-8pt]
     \scalebox{0.98}{Synthesis method} & \scalebox{0.98}{Onomatopoeic word} & \scalebox{0.98}{Sound event label} & \scalebox{0.98}{Method I-1} & \scalebox{0.98}{Method I-2} & \scalebox{0.98}{Method II-1} & \scalebox{0.98}{Method II-2} &\\
    \wcline{1-8}
     &\\[-8pt]
    \scalebox{0.98}{WaveNet} \cite{okamoto_arXiv_2019_01}&  & \Checkmark & \Checkmark & \Checkmark & & &\Checkmark\\
    \cline{1-8}
     &\\[-8pt]
     \scalebox{0.98}{Seq2seq} \cite{okamoto_ATSIP_2022_01}& \Checkmark & & & & \Checkmark & \Checkmark &\\
    \cline{1-8}
     &\\[-8pt]
     \scalebox{0.98}{Seq2seq + event label} \cite{okamoto_ATSIP_2022_01} & \Checkmark & \Checkmark & \Checkmark & \Checkmark & \Checkmark & \Checkmark &\Checkmark\\
    \cline{1-8}
     &\\[-8pt]
     \scalebox{0.98}{Transformer} & \Checkmark & & & & \Checkmark &  \Checkmark &\\
    \cline{1-8}
     &\\[-8pt]
     \scalebox{0.98}{Transformer + event label} & \Checkmark & \Checkmark & \Checkmark &  \Checkmark & \Checkmark &  \Checkmark & \Checkmark\\
    \wcline{1-8}
\end{tabular}
\end{table*}
%

\subsection{Conventional Subjective Evaluation Method}
Subjective evaluation methods have been proposed to subjectively evaluate the quality of synthesized environmental sounds \cite{okamoto_arXiv_2019_01,okamoto_ATSIP_2022_01}.
In this evaluation, the listeners listen to a sound and assign a five-scale absolute score.
However, unlike speech sounds, nonspeech sounds, such as ``{\it a sound of grinding beans in a coffee mill},'' tend to be scored lower even if the sound is of high quality. 
Also, assigning the absolute score of nonspeech sounds is difficult for human listeners, and we cannot obtain consistent answers among human listeners \cite{okamoto_arXiv_2019_01}.
In addition, the evaluation does not indicate whether the synthesized sound reflects the input information (a sound event label in this case).
The synthesized sound is presented to human listeners in the evaluation, but the input information used to synthesize it is not.
Therefore, it is necessary to subjectively evaluate the synthesized sound not only in terms of its quality but also in terms of its degree of conceptual alignment with the input information.

A subjective evaluation of whether the synthesized sounds reflect the input information has also been conducted by Okamoto et al. \cite{okamoto_ATSIP_2022_01}.
An environmental sound and the input information (an onomatopoeic word in this case) are presented to listeners, and the listeners assign a five-scale absolute score.
Their evaluation method solves one of the aforementioned issues; it measures the conceptual alignment between the sound and the input information.
However, the method is limited to absolute evaluation and still suffers from the aforementioned inadequacy of absolute evaluation.
%
%
%
\section{Proposed subjective evaluation methodology}
\label{SecIII}
In this section, we propose a subjective evaluation methodology to evaluate whether the synthesized sound reflects the input information.
To discuss the aforementioned inadequacy of absolute evaluation, our proposal methodology includes both absolute and relative evaluations.
We discuss how subjective evaluations should be conducted by comparing the results of these evaluations.
The sounds to be evaluated in this study are synthesized from either (1) only sound event labels, (2) only onomatopoeic words, or (3) both sound event labels and onomatopoeic words, as input information candidates.
In this work, we evaluate the appropriateness of the sound relative to the input information as the evaluation metric of ``appropriateness.''
In the following evaluation, we propose a methodology to evaluate environmental sounds from the aspect of each input information.

\subsection{Subjective Evaluation Method for Synthesized Sound Using Sound Event Labels}
We propose two types of subjective evaluation for environmental sounds synthesized from sound event labels.

\begin{itemize}
    \item {\bf Evaluation method I-1: Absolute scoring of sound appropriateness to sound event label}\\
    We present a sound and sound event label to listeners.
    The listeners score the appropriateness of the sound for a given sound event label on a scale of 1 (not very appropriate) to 5 (very appropriate).\\

    \item{\bf Evaluation method I-2: Relative preference of sound appropriateness to sound event label }\\
    We present a pair of sounds and a sound event label to listeners. 
    The listeners select the sound perceived to be more appropriate to the label.\\
\end{itemize}

\subsection{Subjective Evaluation Method for Synthesized Sound Using Onomatopoeic Words}
We propose two types of subjective evaluation for environmental sounds synthesized from an onomatopoeic word.

\begin{itemize}
    \item {\bf Evaluation method II-1: Absolute scoring of sound appropriateness to onomatopoeic word}\\
    We present a sound and an onomatopoeic word to listeners.
    The listeners score the appropriateness of the sound for a given onomatopoeic word on a scale of 1 (not very appropriate) to 5 (very appropriate).\\

    \item{\bf Evaluation method II-2: Relative preference of sound appropriateness to onomatopoeic word}\\
    We present a pair of sounds and an onomatopoeic word to listeners.
    The listeners select the sound perceived to be more appropriate to the onomatopoeic word.
\end{itemize}

\begin{table}[t!]
\caption{Information presented to listeners in each subjective evaluation}
\label{table:presented_information}
\centering
\begin{tabular}{lccc}
    \wcline{1-4}
    &\\[-8pt]
    \multirow{2}{*}{\scalebox{0.9}{Method}} & \multirow{2}{*}{\scalebox{0.9}{Sound}} & \scalebox{0.9}{Sound event} & \scalebox{0.9}{Onomatopoeic} \\
     &  & label & word \\
    \wcline{1-4}
    &\\[-10pt]
    \scalebox{0.9}{Okamoto et al. \cite{okamoto_arXiv_2019_01}} & \scalebox{0.9}{single} & &\\
    \cline{1-4}
    \scalebox{0.9}{Liu et al. \cite{Liu_arXiv_2020_01}} & \scalebox{0.9}{single} & & \\
    \cline{1-4}
    \scalebox{0.9}{Yang et al. \cite{Yang_arXiv2022_01}} & \scalebox{0.9}{single} & & \\
    \cline{1-4}
    \scalebox{0.9}{Okamoto et al. \cite{okamoto_ATSIP_2022_01}} & \scalebox{0.9}{single} & & \Checkmark \\
    \cline{1-4}
    \scalebox{0.9}{Method I-1 {\bf (ours)}} & \scalebox{0.9}{single} & \Checkmark & \\
    \cline{1-4}
    \scalebox{0.9}{Method I-2 {\bf (ours)}} & \scalebox{0.9}{pair} & \Checkmark & \\
    \cline{1-4}
    \scalebox{0.9}{Method II-1 {\bf (ours)}} & \scalebox{0.9}{single} & & \Checkmark \\
    \cline{1-4}
    \scalebox{0.9}{Method II-2 {\bf (ours)}} & \scalebox{0.9}{pair} & &\Checkmark \\
    \\[-12pt]
    \wcline{1-4}
\end{tabular}
\end{table}
%
%
\section{Experiments}
\label{SecIV}
In this section, we present results of subjective and objective evaluations.
Moreover, we compare the results of the proposed subjective and conventional objective evaluations for synthesized sounds.
The environmental sound synthesis methods used in the experiments are shown in Table \ref{table:synthesize_method}.
We used the crowdsourcing service to conduct the subjective evaluations.
For method I-1, 100 listeners participated in each evaluation, and each listener evaluated 25 audio samples; for method I-2, 150 listeners participated in each evaluation, and each listener evaluated 25 audio pairs; for method II-1, 100 listeners participated in each evaluation, and each listener evaluated 30 audio samples; for method II-2, 300 listeners participated in each evaluation, and each listener evaluated 25 audio pairs.
The information presented to listeners for each subjective evaluation is shown in Table \ref{table:presented_information}.
%

\subsection{Experimental Conditions}
For model training for environmental sound synthesis, we used 10 types of sound event (bell ringing, alarm cock, manual coffee grinder, cup clinking, drum, maracas, electric shaver, tearing paper, trash box banging, and whistle) included in the Real World Computing Partnership-Sound Scene Database (RWCP-SSD) \cite{Nakamura_AST_1999}.
We used a total of 1,000 audio samples (100 samples $\times$ 10 sound events) in which 95 audio samples of each sound event were used for model training and the remaining five audio samples were used for evaluation.
For onomatopoeic words corresponding to audio samples, we used the RWCP-SSD-Onomatopoeia dataset \cite{okamoto_DCASE_2020}. 
For each audio sample, we used 15 onomatopoeic words, a total of 14,250 onomatopoeic words: (15 onomatopoeic words $\times$ 950 audio samples) from RWCP-SSD-Onomatopoeia.
The hyperparameters for environmental sound synthesis are shown in Table \ref{table:experiment}.

To conduct the objective evaluation introduced in Sec. II-A, we constructed a sound event classifier.
We trained the classifier with the audio samples used for training the environmental sound synthesis model.
The neural network architecture of the sound event classifier was constructed from a three-layer convolutional neural network (CNN) and two-layer fully connected neural network (FC).
The hyperparameters for the sound event classifier are shown in Table \ref{table:experiment_03}.

\begin{table}[t!]
\caption{Experimental conditions for environmental sound synthesis}
\label{table:experiment}
\centering
\begin{tabular}{ll}
    \wcline{1-2}
    &\\[-6pt]
    Sound length & 1--2 s\\
    Sampling rate & 16,000 Hz\\
    Waveform encoding & 16-bit linear PCM\\
    \hline
    &\\[-6pt]
    Acoustic feature & Log-mel \\
    & spectrogram (80 dim.)\\
    Window length for FFT & 0.128 s (2,048 samples) \\
    Window shift for FFT & 0.032 s (512 samples) \\
    \hline
    &\\[-8pt]
    \multicolumn{2}{l}{{\bf Seq2seq / Seq2seq + event label}} \\[2pt]
    \quad Encoder LSTM layers & 1 \\
    \quad \# units in encoder LSTM & 512 \\
    \quad Decoder LSTM layers & 2 \\
    \quad \# units in decoder LSTM & 512, 512 \\
    \quad Event label dimensions & 10 \\
    \quad Teacher forcing rate & 0.6\\
    \quad Batch size & 5 \\
    \quad Optimizer  & RAdam \cite{Liu_ICLR_2020}\\
    \cline{1-2}
    &\\[-8pt]
    \multicolumn{2}{l}{{\bf Transformer / Transformer + event label}} \\[2pt]
    \quad Encoder layers & 3 \\
    \quad Decoder layers & 3 \\
    \quad \# Multi-head & 4 \\
    \quad Batch size & 32 \\
    \quad Event label dimensions & 10 \\
    \quad Optimizer  & RAdam \\
    \wcline{1-2}
\end{tabular}
\end{table}
\begin{table}[t!]
\caption{Experimental conditions for sound event classifier}
\label{table:experiment_03}
\centering
\begin{tabular}{ll}
    \wcline{1-2}
    &\\[-6pt]
    Acoustic feature & Log-mel \\
    & spectrogram (64 dim.)\\
    Window length for FFT & 0.04 s (640 samples)\\
    Window shift for FFT & 0.02 s (320 samples)\\
    \cline{1-2}
    CNN layers & 3 \\
    \# channels of CNN layers & 32, 64, 64 \\
    Filter size & 3$\times$5\\
    Pooling size & 3$\times$3, 3$\times$3 (max pooling)\\
    FC layers & 2 \\
    \# units in FC layer & 64, 128 \\
    \wcline{1-2}
\end{tabular}
\end{table}
\begin{figure*}[t!]
\centering
\includegraphics[scale=0.90]{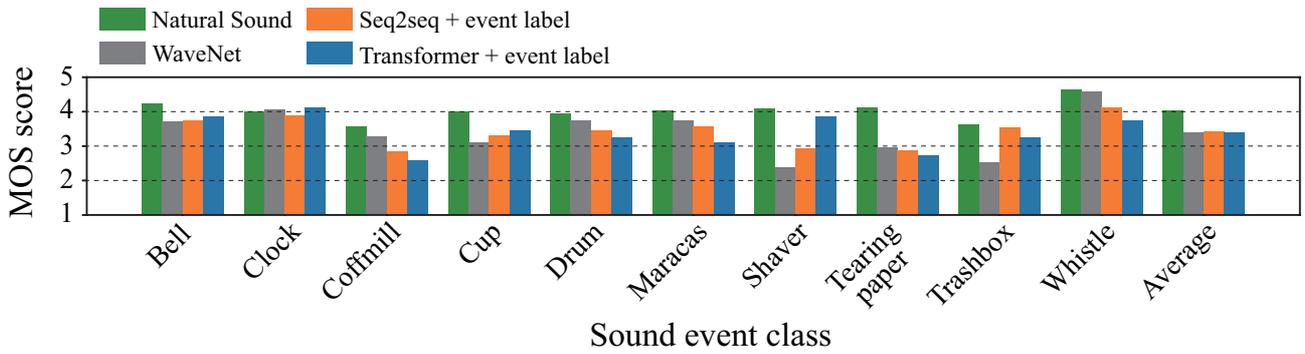}
\vspace{-5pt}
\caption{Absolute evaluation results of the appropriateness of environmental sounds to sound event labels}
\label{fig:methodI_1}
\end{figure*}
\begin{figure}[t!]
\centering
\includegraphics[scale=0.90]{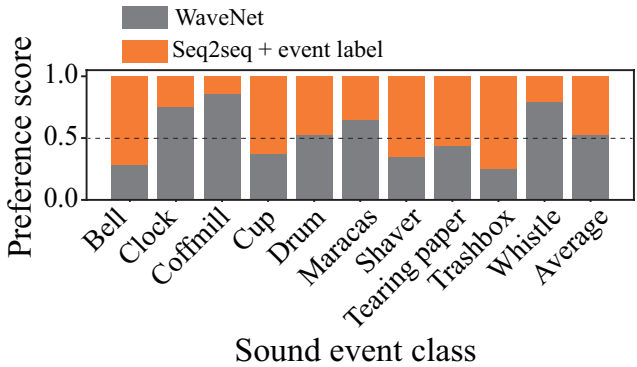}
\vspace{-5pt}
\caption{Relative evaluation results of the appropriateness of environmental sounds to sound event labels (WaveNet vs Seq2seq + event label)}
\label{fig:methodI_2_1}
\end{figure}
\begin{figure}[t!]
\centering
\includegraphics[scale=0.90]{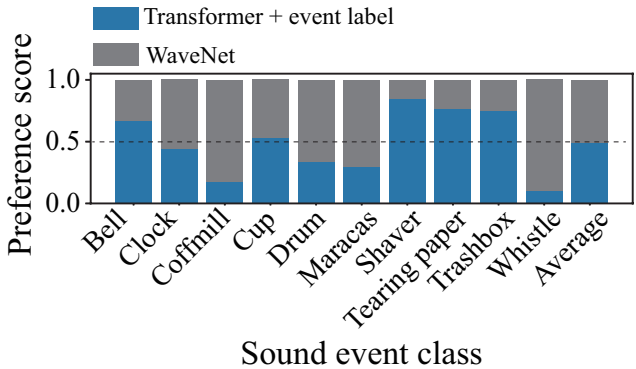}
\vspace{-5pt}
\caption{Relative evaluation results of the appropriateness of environmental sounds to sound event labels (Transformer + event label vs WaveNet)}
\label{fig:methodI_2_2}
\end{figure}
\begin{figure}[t!]
\centering
\includegraphics[scale=0.90]{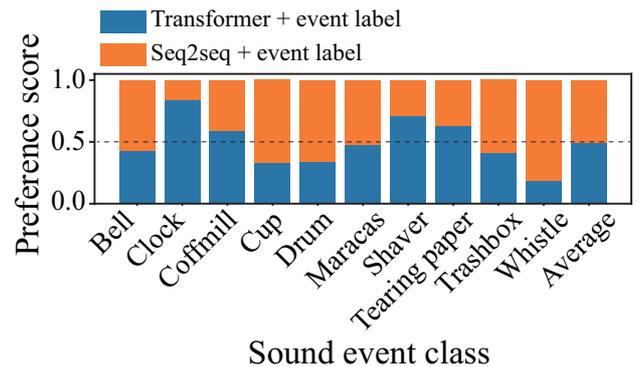}
\vspace{-5pt}
\caption{Relative evaluation results of the appropriateness of environmental sounds to sound event labels (Transformer + event label vs Seq2seq + event label)}
\label{fig:methodI_2_3}
\end{figure}
\begin{figure}[t!]
\centering
\includegraphics[scale=0.70]{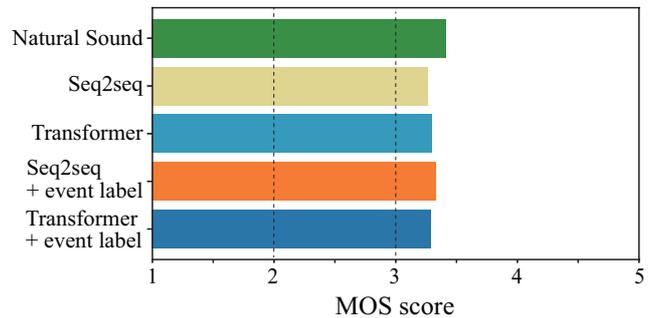}
\vspace{-5pt}
\caption{Absolute evaluation results of the appropriateness of environmental sounds to onomatopoeic words}
\label{fig:methodII_1}
\end{figure}
\begin{figure*}[t!]
\centering
\includegraphics[scale=0.85]{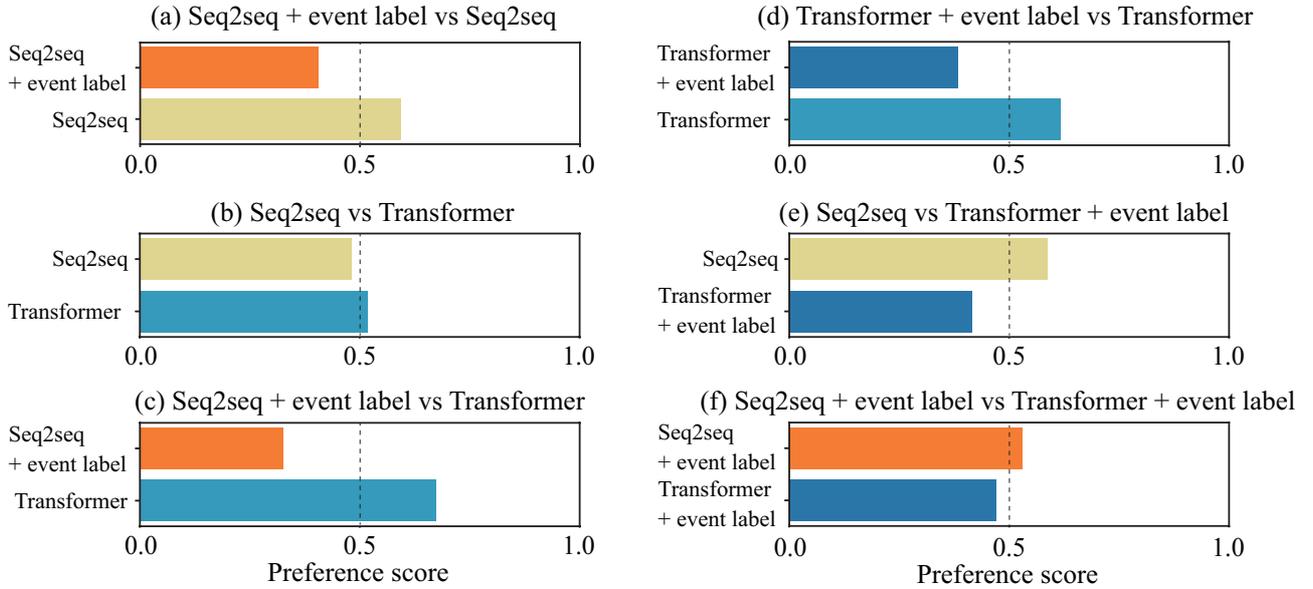}
\vspace{-5pt}
\caption{Relative evaluation results of the appropriateness of environmental sounds to onomatopoeic words}
\label{fig:methodII_2}
\end{figure*}
\begin{figure*}[t!]
\centering
\includegraphics[scale=0.98]{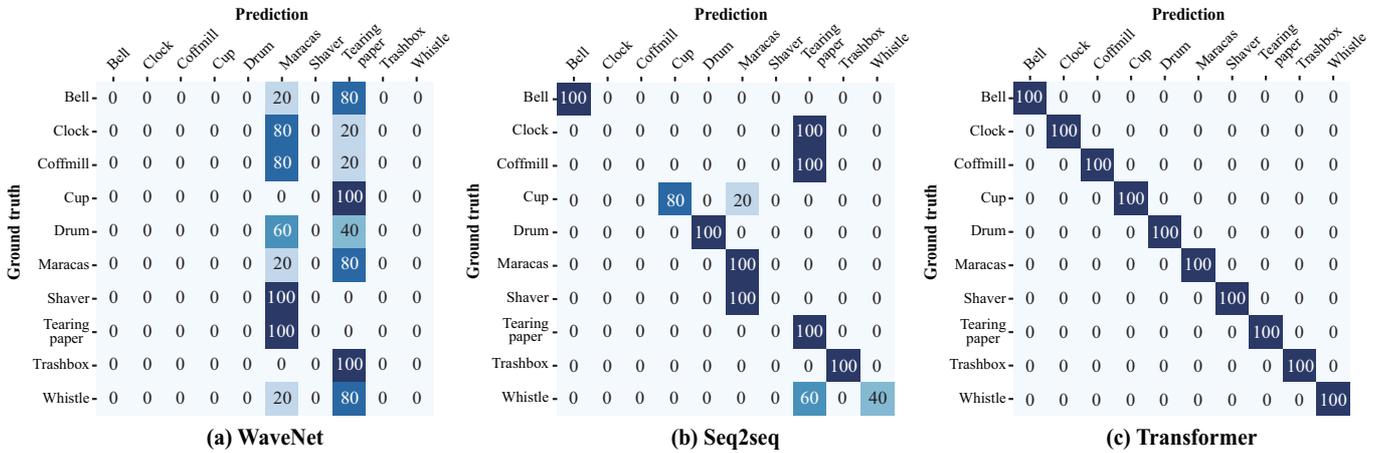}
\caption{Confusion matrix of classification accuracy for synthesized sounds in terms of recall}
\label{fig:objective_WaveNet}
\end{figure*}
\begin{figure}[t!]
\centering
\includegraphics[scale=0.89]{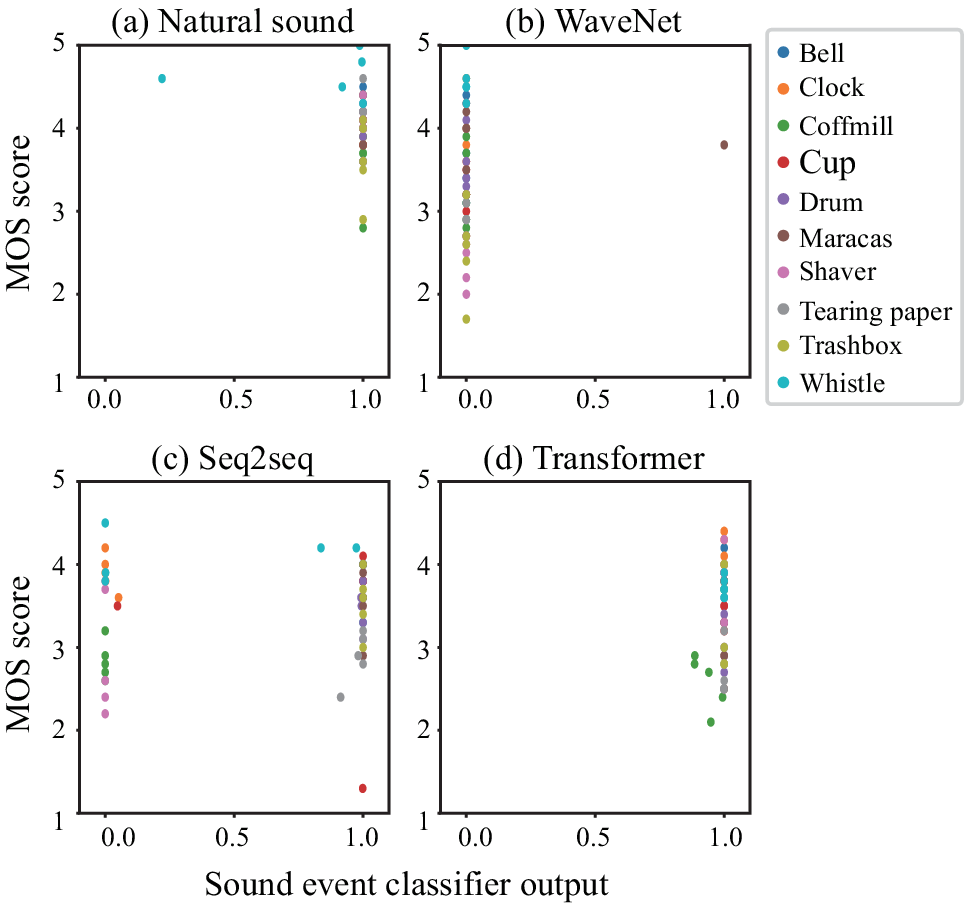}
\vspace{-5pt}
\caption{Scatter plot of sound event classifier output and MOS score.}
\label{figure:expIII}
\end{figure}
\subsection{Results}
\subsubsection{Result of subjective evaluation of synthesized sound using sound event labels}
Figure \ref{fig:methodI_1} shows the results of evaluation method I-1.
The figure shows that there is no marked difference between the synthesis methods in terms of the overall average score (average).
These results indicate that each synthesis method can generate sounds that represent the input sound event labels with the same quality.

Figures \ref{fig:methodI_2_1}, \ref{fig:methodI_2_2}, and \ref{fig:methodI_2_3} show the results of evaluation method I-2.
The figure shows that there was no marked difference between the average scores of the methods, similarly to the results of absolute of evaluation method I-1.
Focusing on the ``sound of maracas,'' the method using the seq2seq + event label obtained a higher score than the method using the Transformer + event label in the absolute evaluation of Fig.~\ref{fig:methodI_1}.
However, in Fig.~\ref{fig:methodI_2_3}, the method using the seq2seq + event label and the method using the Transformer + event label have similar scores (p-value = 0.76).
These results indicate that not only absolute evaluation but also relative evaluation is necessary in the evaluation of synthesized environmental sounds using sound event labels.

\subsubsection{Result of subjective evaluation of synthesized sound using onomatopoeic words}
Figure \ref{fig:methodII_1} shows that the results of evaluation method II-1.
The figure shows that there is no marked difference in the appropriateness of the synthesized sounds for onomatopoeic words among the synthesis methods.
Therefore, from the results of the absolute evaluation, it can be confirmed that whether onomatopoeic words as well as sound event labels are used as input during synthesis, they have no effect on the quality of the synthesized sounds.

Figure \ref{fig:methodII_2} shows the results of evaluation method II-2.
Figures \ref{fig:methodII_2} (a) and (d) show that the synthesis method using only onomatopoeic words as input is more appropriate for onomatopoeic words than the synthesis method using both onomatopoeic words and sound event labels as input.
Since these methods were of similar quality in the absolute evaluation of Method II-1, it was found necessary to conduct not only absolute evaluation but also relative evaluation of synthesized environmental sounds using onomatopoeic words.

\subsubsection{Result of objective evaluation of synthesized sound using sound event labels}
Figure \ref{fig:objective_WaveNet} shows the results of objective evaluation using the sound event classifier.
Figure \ref{fig:objective_WaveNet} (a) shows that most of the sounds synthesized using WaveNet are not classified into the correct class by the sound event classifier.
On the other hand, the results in Figs.~\ref{fig:objective_WaveNet} (b) and (c) show that the sounds synthesized using seq2seq and Transformer can be classified into the correct class by the sound event classifier.

\subsubsection{Comparison of results of subjective and objective evaluations}
Figure \ref{figure:expIII} shows the results of comparing the results of subjective and objective evaluations.
These results show that even when the event classifier classifies the synthesized sound into the correct sound event class, the subjective evaluation score is not high.
On the other hand, when the objective evaluation does not classify the synthesized sound into the correct event class, there are some synthesized sounds that obtain high scores in the subjective evaluation.
An objective evaluation using a classifier classifies sounds similar to the training data into the correct class with higher accuracy.
Thus, for example, if a generated sound that reflects the sound event labels input to the system and its characteristics differ from those of the training data, it tends not to be classified into the correct class.
The evaluation of synthesized environmental sounds used for human listening, such as media content, requires not only objective evaluation but also subjective evaluation.
\section{Conclusions}
\label{SecV}
In this paper, we discussed how to evaluate synthesized environmental sounds.
We proposed a subjective evaluation methodology and compared the results of subjective and objective evaluations.
The experimental results indicated that the results of subjective evaluations tended to differ from those of objective evaluations.
Therefore, it is necessary to conduct not only objective evaluations but also subjective evaluations to evaluate synthesized environmental sounds.
In the future, we will investigate an objective evaluation method for environmental sound synthesis using onomatopoeic words and compare its results with those of subjective evaluation.

\section*{Acknowledgment}

This work was supported by JST SPRING, Grant Number JPMJSP2101, JSPS KAKENHI Grants Number 22J14091 and 22H03639, and ROIS NII Open Collaborative Research 2022-(22S0503).

\bibliographystyle{IEEEtran}
\bibliography{refs}

\end{document}